\documentclass[nofootinbib,prd,showpacs,twocolumn]{revtex4}%
\usepackage{amsmath}
\usepackage{amsfonts}
\usepackage{amssymb}
\usepackage{graphicx}%
\begin{document}
\title{Self-sustained traversable wormholes in noncommutative geometry}
\author{Remo Garattini}
\email{Remo.Garattini@unibg.it}
\affiliation{Universit\`{a} degli Studi di Bergamo, Facolt\`{a} di Ingegneria, Viale
Marconi 5, 24044 Dalmine (Bergamo) ITALY}
\affiliation{INFN - sezione di Milano, Via Celoria 16, Milan, Italy.}
\author{Francisco S. N. Lobo}
\email{flobo@cosmo.fis.fc.ul.pt}
\affiliation{Institute of Gravitation \& Cosmology, University of Portsmouth, Portsmouth
PO1 2EG, UK}
\affiliation{Centro de Astronomia e Astrof\'{\i}sica da Universidade de Lisboa, Campo
Grande, Ed. C8 1749-016 Lisboa, Portugal}
\date{\today}

\begin{abstract}
In this work, we find exact wormhole solutions in the context of
noncommutative geometry, and further explore their physical
properties and characteristics. The energy density of these
wormhole geometries is a smeared and particle-like gravitational
source, where the mass is diffused throughout a region of linear
dimension $\sqrt{\alpha}$ due to the intrinsic uncertainty encoded
in the coordinate commutator. Furthermore, we also analyze these
wormhole geometries considering that the equation governing
quantum fluctuations behaves as a backreaction equation. In
particular, the energy density of the graviton one loop
contribution to a classical energy in a traversable wormhole
background and the finite one loop energy density is considered as
a self-consistent source for these wormhole geometries.
Interesting solutions are found for an appropriate range of the
parameters, validating the perturbative computation introduced in
this semi-classical approach.

\end{abstract}

\maketitle

\section{Introduction}

An interesting development of string/M-theory has been the necessity for
spacetime quantization, where the spacetime coordinates become noncommuting
operators on a $D$-brane \cite{Witten}. The noncommutativity of spacetime is
encoded in the commutator $\left[  \mathbf{x}^{\mu},\mathbf{x}^{\nu}\right]
=i\,\theta^{\mu\nu}$, where $\theta^{\mu\nu}$ is an antisymmetric matrix which
determines the fundamental discretization of spacetime. It has also been shown
that noncommutativity eliminates point-like structures in favor of smeared
objects in flat spacetime \cite{Smailagic:2003yb}. Thus, one may consider the
possibility that noncommutativity could cure the divergences that appear in
general relativity. The effect of the smearing is mathematically implemented
with a substitution of the Dirac-delta function by a Gaussian distribution of
minimal length $\sqrt{\alpha}$. In particular, the energy density of a static
and spherically symmetric, smeared and particle-like gravitational source has
been considered in the following form \cite{Nicolini:2005vd}%
\begin{equation}
\rho_{\alpha}(r)=\frac{M}{(4\pi\alpha)^{3/2}}\;\mathrm{exp}\left(
-\frac{r^{2}}{4\alpha}\right)  \,, \label{NCGenergy}%
\end{equation}
where the mass $M$ is diffused throughout a region of linear dimension
$\sqrt{\alpha}$ due to the intrinsic uncertainty encoded in the coordinate commutator.

In this context, the Schwarzschild metric is modified when a non-commutative
spacetime is taken into account \cite{Nicolini:2005vd, Esposito}. The solution
obtained is described by the following spacetime metric%
\begin{equation}
ds^{2}=-f(r)\,dt^{2}+\frac{dr^{2}}{f(r)}+r^{2}\,(d\theta^{2}+\sin^{2}{\theta
}\,d\phi^{2})\,,\label{NCW}%
\end{equation}
where the factor $f(r)$ is defined as%
\begin{equation}
f(r)=\left[  1-\frac{2r_{S}}{r\sqrt{\pi}}\gamma\left(  \frac{3}{2},\frac
{r^{2}}{4\alpha}\right)  \right]  \,,
\end{equation}
and%
\begin{equation}
\gamma\left(  \frac{3}{2},\frac{r^{2}}{4\alpha}\right)  =\int\limits_{0}%
^{r^{2}/4\alpha}dt\sqrt{t}\exp\left(  -t\right)
\end{equation}
is the lower incomplete gamma function, $r_{S}=2MG$ is the
Schwarzschild radius \cite{Nicolini:2005vd}. The classical
Schwarzschild mass is recovered in the limit
$r/\sqrt{\alpha}\rightarrow\infty$. It was shown that the
coordinate noncommutativity cures the usual problems encountered
in the description of the terminal phase of black hole
evaporation. More specifically, it was found that the evaporation
end-point is a zero temperature extremal black hole and there
exist a finite maximum temperature that a black hole can reach
before cooling down to absolute zero. The existence of a regular
de Sitter at the origin's neighborhood was also shown, implying
the absence of a curvature singularity at the origin. Recently,
further research on noncommutative black holes has been
undertaken, with new solutions found providing smeared source
terms for charged and higher dimensional cases \cite{newNCGbh}.

In this paper we extend the above analysis to wormhole geometries. In
classical general relativity, wormholes are supported by exotic matter, which
involves a stress energy tensor that violates the null energy condition (NEC)
\cite{Morris:1988cz}. Several candidates have been proposed in the literature,
amongst which we refer to the first solutions with what we now call massless
phantom scalar fields \cite{Ellis-Bronn}; solutions in higher dimensions, for
instance in Einstein-Gauss-Bonnet theory \cite{EGB}; wormholes on the brane
\cite{braneWH}; solutions in Brans-Dicke theory \cite{Nandi:1997en}; wormhole
solutions in semi-classical gravity (see Ref. \cite{Garattini:2007ff} and
references therein); exact wormhole solutions using a more systematic
geometric approach were found \cite{Boehmer:2007rm}; solutions supported by
equations of state responsible for the cosmic acceleration \cite{phantomWH};
and NEC respecting geometries were further explored in conformal Weyl gravity
\cite{Lobo:2008zu}, etc (see Refs. \cite{Lemos:2003jb,Lobo:2007zb} for more
details and \cite{Lobo:2007zb} for a recent review).

In this context, we shall also be interested in the analysis that these
wormholes be sustained by their own quantum fluctuations. As wormholes violate
the NEC, and consequently violate all of the other classical energy
conditions, it seems that these exotic spacetimes arise naturally in the
quantum regime, as a large number of quantum systems have been shown to
violate the energy conditions, such as the Casimir effect. Indeed, various
wormhole solutions in semi-classical gravity have been considered in the
literature \cite{semiclassWH}. In this work, we consider the formalism
outlined in detail in Ref. \cite{Garattini,Garattini:2007ff}, where the
graviton one loop contribution to a classical energy in a wormhole background
is used. The latter contribution is evaluated through a variational approach
with Gaussian trial wave functionals, and the divergences are treated with a
zeta function regularization. Using a renormalization procedure, the finite
one loop energy was considered a self-consistent source for a traversable wormhole.

This paper is outlined in the following manner: In Section \ref{sec:II}, we
consider traversable wormholes in noncommutative geometry, by analyzing the
field equations and the characteristics and properties of the shape function
in detail. In Section \ref{sec:III}, we analyze these wormhole geometries in
semi-classical gravity, considering that the equation governing quantum
fluctuations behaves as a backreaction equation. Finally, in Section
\ref{sec:conclusion}, we conclude.

\section{Traversable wormholes in noncommutative geometry}

\label{sec:II}

\subsection{Metric and field equations}

Consider the following static and spherically symmetric spacetime metric%
\begin{equation}
ds^{2}=-e^{2\Phi(r)}\,dt^{2}+\frac{dr^{2}}{1-b(r)/r}+r^{2}\,(d\theta^{2}%
+\sin^{2}{\theta}\,d\phi^{2})\,, \label{metricwormhole}%
\end{equation}
which describes a wormhole geometry with two identical, asymptotically flat
regions joined together at the throat $r_{0}>0$. $\Phi(r)$ and $b(r)$ are
arbitrary functions of the radial coordinate $r$, denoted as the redshift
function and the shape function, respectively. The radial coordinate has a
range that increases from a minimum value at $r_{0}$, corresponding to the
wormhole throat, to $\infty$.

Using the Einstein field equation, $G_{\mu\nu}=8\pi G\,T_{\mu\nu}$ (with
$c=1$), we obtain the following stress-energy tensor scenario%
\begin{align}
\rho(r)  &  =\frac{1}{8\pi G}\;\frac{b^{\prime}}{r^{2}}\,,\label{rhoWH}\\
p_{r}(r)  &  =\frac{1}{8\pi G}\left[  2\left(  1-\frac{b}{r}\right)
\frac{\Phi^{\prime}}{r}-\frac{b}{r^{3}}\right]  \,,\label{prWH}\\
p_{t}(r)  &  =\frac{1}{8\pi G}\left(  1-\frac{b}{r}\right)  \Bigg[\Phi
^{\prime\prime}+(\Phi^{\prime})^{2}\nonumber\\
&  -\frac{b^{\prime}r-b}{2r(r-b)}\Phi^{\prime}-\frac{b^{\prime}r-b}%
{2r^{2}(r-b)}+\frac{\Phi^{\prime}}{r}\Bigg]\,, \label{ptWH}%
\end{align}
in which $\rho(r)$ is the energy density, $p_{r}(r)$ is the radial pressure,
and $p_{t}(r)$ is the lateral pressure measured in the orthogonal direction to
the radial direction. Using the conservation of the stress-energy tensor,
$T^{\mu\nu}{}_{;\nu}=0$, we obtain the following equation%
\begin{equation}
p_{r}^{\prime}=\frac{2}{r}\,(p_{t}-p_{r})-(\rho+p_{r})\,\Phi^{\prime}\,.
\label{prderivative}%
\end{equation}
Note that Eq. $\left(  \ref{prderivative}\right)  $ can also be obtained from
the field equations by eliminating the term $\Phi^{\prime\prime}$ and taking
into account the radial derivative of eq. $\left(  \ref{prWH}\right)  $.

Another fundamental property of wormholes is the violation of the null energy
condition (NEC), $T_{\mu\nu}k^{\mu}k^{\nu}\geq0$, where $k^{\mu}$ is
\textit{any} null vector \cite{Morris:1988cz}. From Eqs. $\left(
\ref{rhoWH}\right)  $ and $\left(  \ref{prWH}\right)  $, considering an
orthonormal reference frame with $k^{\hat{\mu}}=(1,1,0,0)$, so that
$T_{\hat{\mu}\hat{\nu}}k^{\hat{\mu}}k^{\hat{\nu}}=\rho+p_{r}$, one verifies%
\begin{equation}
\rho(r)+p_{r}(r)=\frac{1}{8\pi G}\,\left[  \frac{b^{\prime}r-b}{r^{3}%
}+2\left(  1-\frac{b}{r}\right)  \frac{\Phi^{\prime}}{r}\right]  \,.
\end{equation}
Evaluated at the throat, $r_{0}$, and considering the flaring out condition
\cite{Morris:1988cz}, given by $(b-b^{\prime}r)/b^{2}>0$, and the finite
character of $\Phi(r)$, we have $\rho+p_{r}<0$. Matter that violates the NEC
is denoted \textit{exotic matter}.

We also write out the curvature scalar, $R$, which shall be used below in the
analysis related to the graviton one loop contribution to a classical energy
in a wormhole background. Thus, using metric (\ref{metricwormhole}), the
curvature scalar is given by%
\begin{align}
R  &  =-2\left(  1-\frac{b}{r}\right)  \Bigg[\Phi^{\prime\prime}+(\Phi
^{\prime})^{2}\nonumber\\
&  -\frac{b^{\prime}}{r(r-b)}-\frac{b^{\prime}r+3b-4r}{2r(r-b)}\,\Phi^{\prime
}\Bigg]\,. \label{Ricciscalar}%
\end{align}
We shall henceforth consider a constant redshift function, $\Phi^{\prime
}(r)=0$, which provides interestingly enough results, so that the curvature
scalar reduces to $R=2b^{\prime}/r^{2}$.

\subsection{Analysis of the shape function}

\label{sec:shape}

Throughout this work, we consider the energy density given by Eq.
(\ref{NCGenergy}). Thus, the $(tt)$-component of the Einstein field equation,
Eq. $\left(  \ref{rhoWH}\right)  $, is immediately integrated, and provides
the following solution%
\begin{equation}
b(r)=C+\frac{8\pi GM}{(4\pi\alpha)^{3/2}}\int_{r_{0}}^{r}\,\bar{r}^{2}%
\exp\left(  -\frac{\bar{r}^{2}}{4\alpha}\right)  \,d\bar{r}\,,
\end{equation}
where $C$ is a constant of integration. To be a wormhole solution we need
$b(r_{0})=r_{0}$ at the throat which imposes the condition $C=r_{0}$.

\begin{figure}[h]
\centering
\includegraphics[width=2.4in]{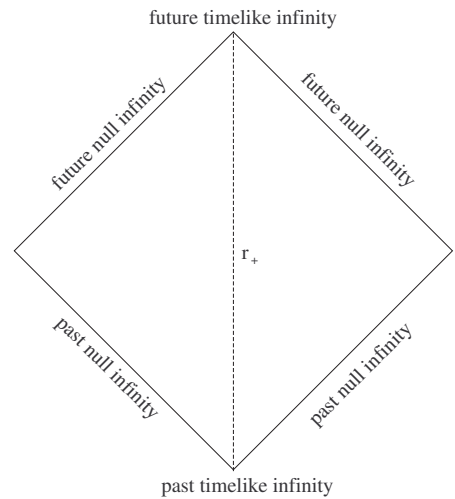}
\caption{Penrose diagram for the asymptotically flat
wormhole spacetime considered in this work.}%
\label{Fig:Penrosediag}%
\end{figure}

However, in the context of noncommutative geometry, without a significant loss
of generality, the above solution may be expressed mathematically in terms of
the incomplete lower gamma functions in the following form%
\begin{equation}
b(r)=C-\frac{2r_{S}}{\sqrt{\pi}}\gamma\left(  \frac{3}{2},\frac{r_{0}^{2}%
}{4\alpha}\right)  +\frac{2r_{S}}{\sqrt{\pi}}\gamma\left(  \frac{3}{2}%
,\frac{r^{2}}{4\alpha}\right)  \,.
\end{equation}
In order to create a correspondence between the spacetime metric (\ref{NCW})
and the spacetime metric of Eq. (\ref{metricwormhole}), we impose that%
\begin{equation}
b(r)=\frac{2r_{S}}{\sqrt{\pi}}\gamma\left(  \frac{3}{2},\frac{r^{2}}{4\alpha
}\right)  \,, \label{NCGform}%
\end{equation}
where%
\begin{equation}
C=\frac{2r_{S}}{\sqrt{\pi}}\gamma\left(  \frac{3}{2},\frac{r_{0}^{2}}{4\alpha
}\right)  \,.
\end{equation}
The shape function expressed in this form is particularly
simplified, as one may now use the properties of the lower
incomplete gamma function, and compare the results directly with
the solution in Ref. \cite{Nicolini:2005vd}. However, note that
the condition $r\geq r_{0}$ is imposed in Eq. $\left(
\ref{NCGform}\right)  $. Thus, throughout this work we consider
the specific case of the shape function given by Eq. $\left(
\ref{NCGform}\right)  $, which provides interesting solutions. See
Fig. \ref{Fig:Penrosediag} for the respective Penrose diagram.

Throughout this work we consider the specific case of a constant redshift
function, i.e., $\Phi^{\prime}=0$, and taking into account the shape function
given by Eq. (\ref{NCGform}), the stress-energy tensor components take the
following form%
\begin{align}
\rho(r)  &  =\frac{r_{S}}{(2\pi\alpha)^{3/2}}\;\mathrm{exp}\left(
-\frac{r^{2}}{4\alpha}\right)  \,,\\
p_{r}(r)  &  =-\frac{r_{S}}{4\pi^{3/2}r^{3}}\;\gamma\left(  \frac{3}{2}%
,\frac{r^{2}}{4\alpha}\right) \\
p_{t}(r)  &  =-\frac{r_{S}}{8\pi^{3/2}r^{3}}\Bigg[\frac{r^{3}}{4\alpha^{3/2}%
}\;\mathrm{exp}\left(  -\frac{r^{2}}{4\alpha}\right)  -\gamma\left(  \frac
{3}{2},\frac{r^{2}}{4\alpha}\right)  \Bigg]
\end{align}

There are some necessary ingredients to be a wormhole solution. First of all,
the existence of a throat expressed by the condition $b(r_{0})=r_{0}$, namely%
\begin{equation}
\frac{1}{a}\gamma\left(  \frac{3}{2},x_{0}^{2}\right)  =x_{0}, \label{throat}%
\end{equation}
where we have defined%
\begin{equation}
x_{0}=\frac{r_{0}}{2\sqrt{\alpha}}\qquad\text{and}\qquad a=\frac{\sqrt
{\alpha\pi}}{r_{S}}\,,
\end{equation}
for notational simplicity.

In order to have one and only one solution we search for extrema of%
\begin{equation}
\frac{1}{x_{0}}\gamma\left(  \frac{3}{2},x_{0}^{2}\right)  =a\,,
\end{equation}
which provides%
\begin{equation}
\bar{x}_{0}=1.5112=\frac{\bar{r}_{0}}{2\sqrt{\alpha}}\,,\quad\text{and}%
\quad\frac{1}{\bar{x}_{0}}\gamma\left(  \frac{3}{2},\bar{x}_{0}^{2}\right)
=0.4654\,. \label{x0}%
\end{equation}
From these relationships we deduce that%
\begin{equation}
\frac{\sqrt{\alpha\pi}}{r_{S}}=0.4654=a\,,
\end{equation}
or%
\begin{equation}
3.81\sqrt{\alpha}=r_{\alpha}=r_{S}\,,
\end{equation}
which finally provides a relationship between $r_{0}$ and $r_{S}$ given by%
\begin{equation}
r_{S}=\bar{r}_{0}\frac{\sqrt{\pi}}{2\bar{x}_{0}a}=1.2599\bar{r}_{0}.
\end{equation}

\begin{figure}[h]
\centering
\includegraphics[width=3.0in]{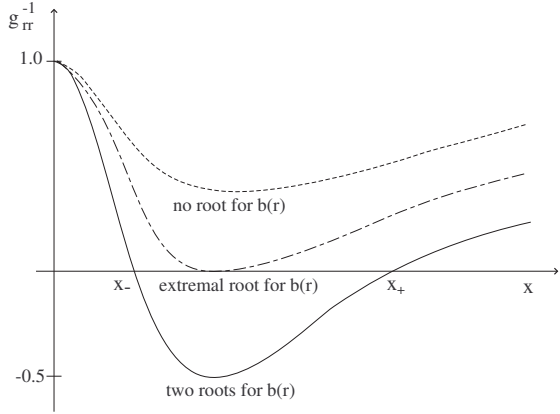}\caption{Plot of $g_{rr}^{-1}$ as a
function of $x=r/\left(  2\sqrt{\alpha}\right)  $ for three
different situations: a) no root, b) extremal root and c) two
roots $x_+=r_+/\left(  2\sqrt{\alpha}\right) $ and $x_-=r_-/\left(
2\sqrt{\alpha}\right)$. See the text for details.}%
\label{Fig:roots}%
\end{figure}

Note that three cases, represented in Fig. \ref{Fig:roots}, need
to be analyzed:

\begin{description}
\item[a)] If $r_{\alpha}>r_{S}$, we have no solutions and therefore no throats;

\item[b)] If $r_{\alpha}<r_{S}$, we have two solutions denoting an inner
throat $r_{-}$ and an outer throat $r_{+}$ with $r_{+}>r_{-}$.

\item[c)] If $r_{\alpha}=r_{S}$, we find the value of Eq. $\left(
\ref{x0}\right)  $. This corresponds to the situation when $r_{+}=r_{-}$ and
it will be interpreted as an \textquotedblleft\textit{extreme}
\textquotedblright\ situation like the extreme Reissner-Nordstr\"{o}m metric.
\end{description}

Unfortunately, case \textbf{c)} does not satisfy the flaring out condition
$b^{\prime}(r_{0})<1$ and therefore will be discarded. We fix our attention to
case \textbf{b)}, which satisfies the flaring out condition. Note that we
necessarily have $r_{+}>\bar{r}_{0}\simeq3\sqrt{\alpha}$. Since the throat
location depends on the value of $a$, we can set without a loss of generality
$r_{+}=k\sqrt{\alpha}$\label{rplus} with $k>3$. Therefore
\begin{equation}
b\left(  r_{+}\right)  =r_{+}=k\sqrt{\alpha}=\frac{2r_{S}}{\sqrt{\pi}}%
\gamma\left(  \frac{3}{2},\frac{k^{2}}{4}\right)  .\label{rkalpha}%
\end{equation}
To avoid the region $r_{-}\leq r\leq r_{+}$ in which $(1-b(r)/r)<0$, we define
the range of $r$ to be $r_{+}\leq r<\infty$. In the analysis below, we require
the form of $b^{\prime}(r_{+})$ and $b^{\prime\prime}(r_{+})$. Thus, by
defining%
\begin{equation}
x=\frac{r_{+}}{2\sqrt{\alpha}},\label{def:x}%
\end{equation}
we deduce the following useful relationships%
\begin{align}
b^{\prime}\left(  r_{+}\right)   &  =\frac{r_{S}r_{+}^{2}}{2\sqrt{\pi
\alpha^{3}}}\exp\left(  -\frac{r_{+}^{2}}{4\alpha}\right)  \nonumber\\
&  =\frac{2x^{3}}{\gamma\left(  \frac{3}{2},x^{2}\right)  }\exp\left(
-x^{2}\right)  \label{db}%
\end{align}
and%
\begin{align}
b^{\prime\prime}\left(  r_{+}\right)   &  =\frac{2r_{S}}{\sqrt{\pi}}%
\exp\left(  -\frac{r_{+}^{2}}{4\alpha}\right)  \frac{r_{+}}{2\sqrt{\alpha^{3}%
}}\left[  1-\frac{r_{+}^{2}}{4\alpha}\right]  \nonumber\\
&  =\frac{2x^{2}}{\sqrt{\alpha}\gamma\left(  \frac{3}{2},x^{2}\right)  }%
\exp\left(  -x^{2}\right)  \left(  1-x^{2}\right)  \,,\label{ddb}%
\end{align}
respectively.

\begin{figure}[h]
\centering
\includegraphics[height=2.0in, width=2.4in]{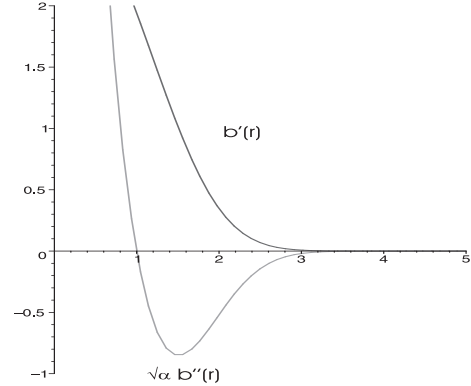}\caption{Plots of
$b^{\prime}(r_{+})$ and $\sqrt{\alpha}b^{\prime\prime}(r_{+})$. Note that
$b^{\prime}\ll1$ for $x\gtrsim3$. This characteristic is analyzed in detail
below, in finding wormhole specific wormhole solutions.}%
\label{Fig:bderivatives}%
\end{figure}

The behavior of $b^{\prime}(r_{+})$ and $b^{\prime\prime}(r_{+})$ are depicted
in Fig. \ref{Fig:bderivatives} for convenience. An extremely interesting case
is that of $b^{\prime}(r_{+})$ $\ll1$, which is valid in the range
$r_{+}\gtrsim6\sqrt{\alpha}$ (or $x\gtrsim3$), as is transparent from the
plots in Fig. \ref{Fig:bderivatives}. This will be explored in detail below.
Note also that $\sqrt{\alpha}b^{\prime\prime}(r_{+})\ll1$ for $x\gtrsim
3$.

\section{Self-sustained wormholes in noncommutative geometry}

\label{sec:III}

In this section, we consider a semi-classical analysis, where the Einstein
field equation takes the form%
\begin{equation}
G_{\mu\nu}=\kappa\,\langle T_{\mu\nu}\rangle^{\mathrm{ren}}\,,
\end{equation}
with $\kappa=8\pi G$, and $\langle T_{\mu\nu}\rangle^{\mathrm{ren}}$ is the
renormalized expectation value of the stress-energy tensor operator of the
quantized field. Now, the metric may be separated into a background component,
$\bar{g}_{\mu\nu}$ and a perturbation $h_{\mu\nu}$, i.e., $g_{\mu\nu}=\bar
{g}_{\mu\nu}+h_{\mu\nu}$. The Einstein tensor may also be separated into a
part describing the curvature due to the background geometry and that due to
the perturbation, i.e.,%
\begin{equation}
G_{\mu\nu}(g_{\alpha\beta})=G_{\mu\nu}(\bar{g}_{\alpha\beta})+\Delta G_{\mu
\nu}(\bar{g}_{\alpha\beta},h_{\alpha\beta})\,,
\end{equation}
where $\Delta G_{\mu\nu}(\bar{g}_{\alpha\beta},h_{\alpha\beta})$ may be
considered a perturbation series in terms of $h_{\mu\nu}$. Using the
semi-classical Einstein field equation, in the absence of matter fields, one
may define an effective stress-energy tensor for the quantum fluctuations as%
\begin{equation}
\langle T_{\mu\nu}\rangle^{\mathrm{ren}}=-\frac{1}{\kappa}\,\langle\Delta
G_{\mu\nu}(\bar{g}_{\alpha\beta})\rangle^{\mathrm{ren}}\,,
\end{equation}
so that the equation governing quantum fluctuations behaves as a backreaction
equation. The semi-classical procedure followed in this work relies heavily on
the formalism outlined in Refs. \cite{Garattini,Garattini:2007ff}, where the
graviton one loop contribution to a classical energy in a traversable wormhole
background was computed, through a variational approach with Gaussian trial
wave functionals \cite{Garattini,Garattini2}. A zeta function regularization
is used to deal with the divergences, and a renormalization procedure is
introduced, where the finite one loop is considered as a self-consistent
source for traversable wormholes. Rather than reproduce the formalism, we
shall refer the reader to Refs. \cite{Garattini,Garattini:2007ff} for details,
when necessary. In this paper, rather than integrate over the whole space as
in Ref. \cite{Garattini}, we shall work with the energy densities, which
provides a more general working hypothesis, which follows the approach
outlined in Ref. \cite{Garattini:2007ff}. However, for self-completeness and
self-consistency, we present here a brief outline of the formalism used.

The classical energy is given by%
\begin{equation}
H_{\Sigma}^{(0)}=\int_{\Sigma}\,d^{3}x\,\mathcal{H}^{(0)}=-\frac{1}{16\pi
G}\int_{\Sigma}\,d^{3}x\,\sqrt{g}\,R\,,
\end{equation}
where the background field super-hamiltonian, $\mathcal{H}^{(0)}$, is
integrated on a constant time hypersurface. $R$ is the curvature scalar given
by Eq. $\left(  \ref{Ricciscalar}\right)  $, and for simplicity we consider
$\Phi^{\prime}=0$, as mentioned above, which provides interesting enough
results. Thus, the classical energy reduces to%
\begin{equation}
H_{\Sigma}^{(0)}=-\frac{1}{2G}\int_{r_{0}}^{\infty}\,\frac{dr\,r^{2}}%
{\sqrt{1-b(r)/r}}\,\frac{b^{\prime}(r)}{r^{2}}\,. \label{classical}%
\end{equation}

We shall also take into account the total regularized one loop energy given by%
\begin{equation}
E^{TT}=2\int_{r_{0}}^{\infty}\,dr\frac{r^{2}}{\sqrt{1-b(r)/r}}\,\left[
\rho_{1}(\varepsilon,\mu)+\rho_{2}(\varepsilon,\mu)\right]  \,,
\end{equation}
where once again, we refer the reader to Refs.
\cite{Garattini,Garattini:2007ff} for details. The energy densities, $\rho
_{i}(\varepsilon)$ (with $i=1,2$), are defined as%
\begin{align}
\rho_{i}(\varepsilon,\mu)  &  =\frac{1}{4\pi}\mu^{2\varepsilon}\int
_{\sqrt{U_{i}(r)}}^{\infty}\,d\tilde{E}_{i}\,\frac{\tilde{E}_{i}^{2}}{\left[
\tilde{E}_{i}^{2}-U_{i}(r)\right]  ^{\varepsilon-1/2}}\nonumber\\
&  =-\frac{U_{i}^{2}(r)}{64\pi^{2}}\left[  \frac{1}{\varepsilon}+\ln\left(
\frac{\mu^{2}}{U_{i}}\right)  +2\ln2-\frac{1}{2}\right]  . \label{energy}%
\end{align}
The zeta function regularization method has been used to determine the energy
densities, $\rho_{i}$. It is interesting to note that this method is identical
to the subtraction procedure of the Casimir energy computation, where the zero
point energy in different backgrounds with the same asymptotic properties is
involved (see also Ref. \cite{DGL} for a related issue in the context of
topology change). In this context, the additional mass parameter $\mu$ has
been introduced to restore the correct dimension for the regularized
quantities. Note that this arbitrary mass scale appears in any regularization scheme.

We emphasize that in analyzing the Einstein field equations, one should
consider the whole system of equations, which in the static spherically
symmetric case also includes the $rr$-component. The joint analysis of the
equations is necessary to guarantee the compatibility of the system. However,
in the semi-classical framework considered in this work there is no dynamical
equation for the pressure. Nevertheless, one may argue that the semi-classical
part of the pressure is known through the equation of state that determines
the relation between the energy density and the pressure.

Since a self sustained wormhole must satisfy%
\begin{equation}
H_{\Sigma}^{(0)}=-\,E^{TT},
\end{equation}
which is an integral relation, this has to be true also for the integrand,
namely the energy density. Therefore, we set%
\begin{equation}
\frac{b^{\prime}(r)}{2Gr^{2}}=2\,\left[  \rho_{1}(\varepsilon,\mu)+\rho
_{2}(\varepsilon,\mu)\right]  \,.\label{Ett}%
\end{equation}
For this purpose, the Lichnerowicz equations provide the potentials, which are
given by%
\begin{equation}
U_{1}(r)=\frac{6}{r^{2}}\,\left[  1-\frac{b(r)}{r}\right]  -\frac{3}{2r^{2}%
}\,\left[  b^{\prime}(r)-\frac{b(r)}{r}\right]  \,,\label{p1}%
\end{equation}%
\begin{equation}
U_{2}(r)=\frac{6}{r^{2}}\,\left[  1-\frac{b(r)}{r}\right]  -\frac{3}{2r^{2}%
}\,\left[  \frac{b^{\prime}(r)}{3}+\frac{b(r)}{r}\right]  .\label{p2}%
\end{equation}
We refer the reader to Refs. \cite{Garattini,Garattini:2007ff} for the
deduction of these expressions. Thus, taking into account Eq. $\left(
\ref{energy}\right)  $, then Eq. $\left(  \ref{Ett}\right)  $ yields the
following relationship%
\begin{align}
\frac{b^{\prime}(r)}{2Gr^{2}} &  =-\frac{1}{32\pi^{2}\varepsilon}\,\left[
U_{1}^{2}(r)+U_{2}^{2}(r)\right]  \nonumber\\
&  \hspace{-1.2cm}-\frac{1}{32\pi^{2}}\left[  U_{1}^{2}\ln\left(  \left\vert
\frac{4\mu^{2}}{U_{1}\sqrt{e}}\right\vert \right)  +U_{2}^{2}\ln\left(
\left\vert \frac{4\mu^{2}}{U_{2}\sqrt{e}}\right\vert \right)  \right]
.\label{selfsust}%
\end{align}
It is essential to renormalize the divergent energy by absorbing the
singularity in the classical quantity, by redefining the bare classical
constant $G$ as%
\begin{equation}
\frac{1}{G}\rightarrow\frac{1}{G_{0}}-\frac{2}{\varepsilon}\,\frac{\left[
U_{1}^{2}(r)+U_{2}^{2}(r)\right]  }{32\pi^{2}}\frac{r^{2}}{b^{\prime}(r)}\,.
\end{equation}
Using this, Eq. $\left(  \ref{selfsust}\right)  $ takes the form%
\begin{equation}
\,\frac{b^{\prime}(r)}{2G_{0}r^{2}}=-\frac{1}{32\pi^{2}}\left[  U_{1}^{2}%
\ln\left(  \left\vert \frac{4\mu^{2}}{U_{1}\sqrt{e}}\right\vert \right)
+U_{2}^{2}\ln\left(  \left\vert \frac{4\mu^{2}}{U_{2}\sqrt{e}}\right\vert
\right)  \right]  .\label{selfsust2}%
\end{equation}
Note that this quantity depends on an arbitrary mass scale. Thus, using the
renormalization group equation to eliminate this dependence, we impose that%
\begin{align}
\mu\frac{d}{d\mu}\left[  \frac{b^{\prime}(r)}{2G_{0}\left(  \mu\right)  r^{2}%
}\right]   &  =-\frac{\mu}{32\pi^{2}}\frac{d}{d\mu}\Bigg[U_{1}^{2}\ln\left(
\left\vert \frac{4\mu^{2}}{U_{1}\sqrt{e}}\right\vert \right)  \nonumber\\
&  +U_{2}^{2}\ln\left(  \left\vert \frac{4\mu^{2}}{U_{2}\sqrt{e}}\right\vert
\right)  \Bigg]\,,
\end{align}
which reduces to%
\begin{equation}
\frac{b^{\prime}(r)}{r^{2}}\,\mu\,\frac{\partial G_{0}^{-1}(\mu)}{\partial\mu
}=-\frac{1}{8\pi^{2}}\,\left[  U_{1}^{2}(r)+U_{2}^{2}(r)\right]
\,.\label{selfsust3}%
\end{equation}
The renormalized constant $G_{0}$ is treated as a running constant, in the
sense that it varies provided that the scale $\mu$ is varying, so that one may
consider the following definition%
\begin{equation}
\frac{1}{G_{0}(\mu)}=\frac{1}{G_{0}(\mu_{0})}-\frac{\left[  U_{1}^{2}%
(r)+U_{2}^{2}(r)\right]  }{8\pi^{2}}\,\,\frac{r^{2}}{b^{\prime}(r)}%
\,\ln\left(  \frac{\mu}{\mu_{0}}\right)  \,,\label{selfsust4a}%
\end{equation}
which can be cast into the following form%
\begin{equation}
G_{0}(\mu)=\frac{G_{0}(\mu_{0})}{1-G_{0}(\mu_{0})a\left(  r\right)  \ln\left(
\frac{\mu}{\mu_{0}}\right)  },
\end{equation}
where
\begin{equation}
a\left(  r\right)
=\frac{\left[  U_{1}^{2}(r)+U_{2}^{2}(r)\right]  }{8\pi^{2}}\,\,\frac{r^{2}%
}{b^{\prime}(r)}.
\end{equation}
We can note that there is a blow up at a scale%
\begin{equation}
\mu_{0}\exp\left(  \frac{1}{G_{0}(\mu_{0})a\left(  r\right)  }\right)
=\mu,\label{Landau}%
\end{equation}
which is very large if the argument of the exponential is large. This is
called a Landau point invalidating the perturbative computation. Thus, Eq.
$\left(  \ref{selfsust}\right)  $ finally provides us with%
\begin{align}
\frac{16\pi^{2}}{G_{0}(\mu_{0})} &  =-\frac{r^{2}}{b^{\prime}(r)}%
\,\Bigg[U_{1}^{2}(r)\ln\left(  \left\vert \frac{4\mu_{0}^{2}}{U_{1}(r)\sqrt
{e}}\right\vert \right)  \nonumber\\
&  +U_{2}^{2}(r)\ln\left(  \left\vert \frac{4\mu_{0}^{2}}{U_{2}(r)\sqrt{e}%
}\right\vert \right)  \Bigg].\label{selfsust4}%
\end{align}
Now, the procedure that we shall follow is to find the extremum of the right
hand side of Eq. $\left(  \ref{selfsust4}\right)  $ with respect to $r$, and
finally evaluate at the throat $r_{0}$, in order to have only one solution
(see discussion in Ref. \cite{Garattini}).

To this effect, we shall use the derivative of the potentials, Eqs. $\left(
\ref{p1}\right)  $-$\left(  \ref{p2}\right)  $, which we write down for
self-completeness and self-consistency, and are given by%
\begin{eqnarray}
U_{1}^{\prime}(r)  &  =&-\frac{12}{r^{3}}\left(
1-\frac{b}{r}\right) +\frac{6(b-b^{\prime}r)}{r^{4}}
   \nonumber   \\
&&-\frac{3}{2r^{2}}\left[  b^{\prime\prime}%
-\frac{b^{\prime}r-b}{r^{2}}\,\right]  ,\\
U_{2}^{\prime}(r)  &  =&-\frac{12}{r^{3}}\left(
1-\frac{b}{r}\right)
+\frac{6(b-b^{\prime}r)}{r^{4}}
    \nonumber   \\
&&-\frac{3}{2r^{2}}\left[ \frac{b^{\prime\prime
}}{3}+\frac{b^{\prime}r-b}{r^{2}}\right]  ,
\end{eqnarray}
and once evaluated at the throat take the following form%
\begin{align}
U_{1}^{\prime}(r_{+})  &  =\frac{9(1-b^{\prime}\left(  r_{+}\right)  )}%
{2r_{+}^{3}}-\frac{3b^{\prime\prime}\left(  r_{+}\right)  }{2r_{+}^{2}%
}\,,\label{p01}\\
U_{2}^{\prime}(r_{+})  &  =\frac{15(1-b^{\prime}\left(  r_{+}\right)
)}{2r_{+}^{3}}-\frac{b^{\prime\prime}\left(  r_{+}\right)  }{2r_{+}^{2}}\,,
\label{p02}%
\end{align}
respectively. The potentials, Eqs. $\left(  \ref{p1}\right)  $-$\left(
\ref{p2}\right)  $ evaluated at the throat reduce to%
\begin{align}
U_{1}(r_{+})  &  =-\frac{3}{2r_{+}^{2}}\left[  b^{\prime}(r_{+})-1\right]
\,,\label{pot1}\\
U_{2}(r_{+})  &  =-\frac{3}{2r_{+}^{2}}\,\left[  \frac{b^{\prime}(r_{+})}%
{3}+1\right]  \,. \label{pot2}%
\end{align}
Thus, the extremum of Eq. $\left(  \ref{selfsust4}\right)  $ with respect to
$r$, takes the form%
\begin{align}
&  \frac{r}{2\alpha}\,\left[  U_{1}^{2}\ln\left(  \frac{|U_{1}|\sqrt{e}}%
{4\mu_{0}^{2}}\right)  +U_{2}^{2}\ln\left(  \frac{|U_{2}|\sqrt{e}}{4\mu
_{0}^{2}}\right)  \right] \nonumber\\
&  \hspace{-0.5cm}+\left[  2U_{1}U_{1}^{\prime}\ln\left(  \frac{|U_{1}|e}%
{4\mu_{0}^{2}}\right)  +2U_{2}U_{2}^{\prime}\ln\left(  \frac{|U_{2}|e}%
{4\mu_{0}^{2}}\right)  \right]  =0, \label{Ap2}%
\end{align}
where the following relationship%
\begin{equation}
\left[  U_{i}^{2}\ln\left(  \frac{|U_{i}|\sqrt{e}}{4\mu_{0}^{2}}\right)
\right]  ^{\prime}=2U_{i}U_{i}^{\prime}\ln\left(  \frac{|U_{i}|e}{4\mu_{0}%
^{2}}\right)  ,
\end{equation}
has been used.

Considering the general features given by Eqs. (\ref{db}) and
(\ref{ddb}) provides an intractable analysis to the problem.
However, an extremely interesting case which we explore in detail
is that of $b^{\prime}(r_{+})$ $\ll1$, which is justified for the
range $r_{+}\gtrsim 6\sqrt{\alpha}$ (or $x\gtrsim3$), as is
transparent from the plots in Fig. \ref{Fig:bderivatives}. Thus,
the potentials, Eqs. $\left(  \ref{p1}\right)
$-$\left(  \ref{p2}\right)  $ evaluated at the throat reduce to%
\begin{align}
U_{1}(r_{+})  &  \simeq\frac{3}{2r_{+}^{2}}\,,\\
U_{2}(r_{+})  &  \simeq-\frac{3}{2r_{+}^{2}}\,,
\end{align}
respectively. The derivatives of the potentials $U_{1}(r_{+})$ and
$U_{2}(r_{+})$ at the throat, Eqs. $\left(  \ref{p01}\right)  $-$\left(
\ref{p02}\right)  $, take the following form%
\begin{align}
U_{1}^{\prime}(r_{+})  &  \simeq\frac{9}{2r_{+}^{3}}-\frac{3b^{\prime\prime
}\left(  r_{+}\right)  }{2r_{+}^{2}}\,,\\
U_{2}^{\prime}(r_{+})  &  \simeq\frac{15}{2r_{+}^{3}}-\frac{b^{\prime\prime
}\left(  r_{+}\right)  }{2r_{+}^{2}}\,.
\end{align}

\begin{figure}[h]
\centering
\includegraphics[height=2.0in, width=2.4in]{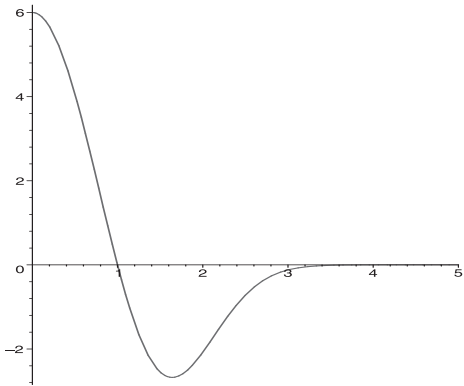}\caption{Plot
depicting the behavior of $r_{+}b^{\prime\prime}(r_{+})$.}%
\label{Fig:rb}%
\end{figure}

Finally, we verify that Eq. $\left(  \ref{Ap2}\right)  $ provides%
\begin{equation}
\frac{3}{4\alpha}\ln\left(  \frac{3\sqrt{e}}{8\mu_{0}^{2}r_{+}^{2}}\right)
-\left[  \frac{6}{r_{+}^{2}}+\frac{b^{\prime\prime}(r_{+})}{r_{+}}\right]
\ln\left(  \frac{3e}{8\mu_{0}^{2}r_{+}^{2}}\right)  =0. \label{approx1}%
\end{equation}
Note that the factor in square brackets in the second term may be expressed as%
\begin{equation}
\left[  \frac{6}{r_{+}^{2}}+\frac{b^{\prime\prime}(r_{+})}{r_{+}}\right]
=\frac{6}{r_{+}^{2}}\left[  1+\frac{1}{6}r_{+}b^{\prime\prime}(r_{+})\right]
\,, \label{bsec}%
\end{equation}
and using the following dimensionless relationship%
\begin{equation}
r_{+}b^{\prime\prime}(r_{+})=\frac{4x^{3}}{\gamma\left(  \frac{3}{2}%
,x^{2}\right)  }\exp\left(  -x^{2}\right)  \left(  1-x^{2}\right)  \,,
\end{equation}
where $x$ is given by Eq. $\left(  \ref{def:x}\right)  $, one immediately
verifies that $r_{+}b^{\prime\prime}(r_{+})\ll1$ for $x\gtrsim3$, which is
transparent from Fig. \ref{Fig:rb}.

Thus, Eq. $\left(  \ref{approx1}\right)  $ reduces to%
\begin{equation}
\ln\left(  \frac{3\sqrt{e}}{8\mu_{0}^{2}r_{+}^{2}}\right)  =\frac{8\alpha
}{r_{+}^{2}}\ln\left(  \frac{3e}{8\mu_{0}^{2}r_{+}^{2}}\right)  \,,
\label{rad}%
\end{equation}
which in turn may be reorganized to yield the following useful relationship%
\begin{equation}
\left(  1-\frac{8}{k^{2}}\right)  \ln\left(  \frac{3e}{8\mu_{0}^{2}\alpha
k^{2}}\right)  =\frac{1}{2}, \label{rad1}%
\end{equation}
where we have set, as defined in Eq. (\ref{rkalpha}), $r_{+}=k\sqrt{\alpha}$,
with $k>4$. Isolating the $k$-dependent term provides the expression%
\begin{equation}
\frac{8\mu_{0}^{2}\alpha}{3\sqrt{e}}=\;k^{-2}\exp\left(  -\frac{4}{k^{2}%
-8}\right)  \,. \label{rad2b}%
\end{equation}
It is interesting to note that the r.h.s. has two extrema for positive $k$:
$k_{1}=2$ and $k_{2}=4$. In terms of the wormhole throat the first value
becomes $r_{1}=2\sqrt{\alpha}$, which will be discarded because its location
is below the extreme radius, as emphasized in Section \ref{sec:shape}.
Relative to the second extrema $k_{2}$, i.e., $r_{2}=4\sqrt{\alpha}$, despite
the fact that $k_{2}$ is larger than the extreme value $\left(  k=3\right)  $,
it does not fall into the range of the approximation of $b^{\prime}\left(
r_{+}\right)  \ll1$. Thus, it may also be discarded. Therefore, we need to
search for values larger than $k=4$. In particular, according to Eq. $\left(
\ref{bsec}\right)  $, to neglect $r_{+}b^{\prime\prime}(r_{+})$ we must set
$k\geq6$. To this purpose, we build the following table%
\begin{equation}%
\begin{tabular}
[c]{|c|c|}\hline
Values of $k$ & $\mu_{0}\sqrt{\alpha}$\\\hline
$6$ & $0.12202$\\\hline
$7$ & $0.10698$\\\hline
$8$ & $0.09484$\\\hline
\end{tabular}
\ \ \ \ . \label{T1}%
\end{equation}
Plugging the expression of $\mu_{0}\sqrt{\alpha}$, i.e., Eq. (\ref{rad2b}),
into Eq.$\left(  \ref{selfsust4}\right)  $, we get%
\begin{equation}
\frac{9\gamma\left(  \frac{3}{2},\frac{k^{2}}{4}\right)  }{2\pi^{2}%
k^{5}\left(  k^{2}-8\right)  }\exp\left(  \frac{k^{2}}{4}\right)
=\,\frac{\alpha}{G_{0}(\mu_{0})}.
\end{equation}
The following table illustrates the behavior of $\sqrt{G_{0}(\mu_{0})/\alpha}%
$, and therefore of $r_{+}/\sqrt{G_{0}(\mu_{0})}$%
\begin{equation}%
\begin{tabular}
[c]{|c|c|c|}\hline
Values of $k$ & $\sqrt{\alpha/G_{0}\left(  \mu_{0}\right)  }$ & $r_{+}%
/\sqrt{G_{0}\left(  \mu_{0}\right)  }$\\\hline
$6$ & $0.12260\allowbreak$ & $\allowbreak\allowbreak0.74$\\\hline
$7$ & $0.35006$ & $2.\,\allowbreak5$\\\hline
$\bar{k}=7.77770$ & $1$ & $\allowbreak\allowbreak7.77770$\\\hline
$8$ & $1.39883$ & $\allowbreak\allowbreak11$\\\hline
$9$ & $7.64175$ & $\allowbreak69.$\\\hline
$10$ & $56.23620$ & $\allowbreak5.\,\allowbreak6\times10^{2}$\\\hline
\end{tabular}
\ \ \ \ . \label{T2}%
\end{equation}
Note that in table $\left(  \ref{T1}\right)  $, as $k$ increases
then $\mu _{0}\sqrt{\alpha}$ decreases. This means that we are
approaching the classical value where the non-commutative
parameter $\alpha\rightarrow0$. It appears also that there exists
a critical value of $k$ where $\alpha=G_{0}\left( \mu_{0}\right)
$. To fix ideas, suppose we fix $G_{0}\left(  \mu_{0}\right) $ at
the Planck scale, then below $\bar{k}$, the non-commutative
parameter becomes more fundamental than $G_{0}\left(
\mu_{0}\right)  $. This could be a signal of another scale
appearing, maybe connected with string theory. On the other hand
above $\bar{k}$, we have the reverse. However, the increasing
values in the table is far to be encouraging because the non
commutative approach breaks down when $k$ is very large.
Nevertheless, note the existence of interesting solutions in the
neighborhood of the value $\bar{k}=7.7$.


\section{Conclusion}

\label{sec:conclusion}

In this work, we have analyzed exact wormhole solutions in the
context of noncommutative geometry. The energy density of these
wormhole geometries is a smeared and particle-like gravitational
source, where the mass is diffused throughout a region of linear
dimension $\sqrt{\alpha}$ due to the intrinsic uncertainty encoded
in the coordinate commutator. The physical properties and
characteristics of these wormhole solutions were further explored.
Finally, we analyzed these wormhole geometries in semi-classical
gravity, considering that the equation governing quantum
fluctuations behaves as a backreaction equation. In particular,
the energy density of the graviton one loop contribution to a
classical energy in a traversable wormhole background and the
finite one loop energy density is considered as a self-consistent
source for these wormhole geometries. What we discover is that
there exists a continuous set of solutions parametrized by $k$.
Apparently, this set of solutions is unbounded for large values of
$k$. However, as the critical value $\bar{k}$ is passed we note a
vary rapidly divergent value of the ratio $\alpha/G_{0}\left(
\mu_{0}\right)  $ with a consequent growing of the wormhole
radius. Nevertheless, this sector of the $k$-range does not
represent physical solutions because we are approaching the region
where $\alpha=0$. Moreover, from Eq. $\left(  \ref{Landau}\right)
$ we can see that the larger the value of $k$, the closer is the
value of the scale $\mu$ to the Landau point or in other words,
for large values of $k$ there is no evolution for the Eq. $\left(
\ref{selfsust4}\right)  $. This is evident by plugging in
expression $\left(  \ref{rad2b}\right)  $ into Eq. $\left(
\ref{selfsust4a} \right)  $ leading to
\begin{equation}
\mu_{0}\exp\left(  \frac{2\alpha}{r^{2}-8\alpha}\right)  =\mu_{0}\exp\left(
\frac{2}{k^{2}-8}\right)  =\mu.
\end{equation}
On the other hand, within the range of validity of the
approximation where $b^{\prime}(r_{+})$ $\ll1$, we find that we
are far from the blow up region which means that the perturbative
computation in this range can be considered correct.


\section*{Acknowledgements}

FSNL was funded by Funda\c{c}\~{a}o para a Ci\^{e}ncia e Tecnologia
(FCT)--Portugal through the research grant SFRH/BPD/26269/2006.

\end{document}